\documentclass[journal, twoside]{IEEEtran}

\usepackage{array}
\usepackage{multirow}
\usepackage{subfigure} 
\usepackage{algorithm}
\usepackage{algorithmic}
\usepackage{dsfont}
\usepackage{graphicx}
\usepackage{textcomp}
\newtheorem{mytheorem}{Theorem}
\usepackage{pseudocode}
\usepackage{amsmath}
\usepackage{amsfonts}
\usepackage{amssymb}

\ifCLASSINFOpdf
  
\else
 
\fi

\begin{document}

\title{A Measurement Framework for Directed Networks}

\author{Mostafa~Salehi 
        and~Hamid~R.~Rabiee
\thanks{The authors are with Department of Computer Engineering, Sharif University of Technology, Tehran, Iran. e-mail: mostafa\_salehi@ce.sharif.edu, rabiee@sharif.edu}
}

\maketitle

\begin{abstract}
Partially-observed network data collected by link-tracing based sampling methods is often being studied to obtain the characteristics of a large complex network. However, little attention has been paid to sampling from directed networks such as WWW and Peer-to-Peer networks. In this paper, we propose a novel two-step (sampling/estimation) framework to measure nodal characteristics which can be defined by an average target function in an arbitrary directed network. To this end, we propose a personalized PageRank-based algorithm to visit and sample nodes. This algorithm only uses already visited nodes as local information without any prior knowledge about the latent structure of the network. Moreover, we introduce a new estimator based on the approximate importance sampling to estimate average target functions. The proposed estimator utilizes calculated PageRank value of each sampled node as an approximation for the exact visiting probability. To the best of our knowledge, this is the first study on correcting the bias of a sampling method by re-weighting of measured values that considers the effect of approximation of visiting probabilities. Comprehensive theoretical and empirical analysis of the estimator demonstrate that it is asymptotically unbiased even in situations where stationary distribution of PageRank is poorly approximated.
\end{abstract}


\IEEEpeerreviewmaketitle

\section{Introduction}
\label{sec:Intro}

\IEEEPARstart{M}{any} real-world communication systems such as the Internet, World Wide Web (WWW), Peer-to-Peer systems, and Online Social Networks (OSNs) can be modeled as a complex network of interacting dynamical nodes. In the last decade, a considerable amount of research has been done on measuring the characteristics of complex networks in various domains \cite{Newman2006}. To extract useful knowledge from a network, one should study the collected network data. However, the tremendous growth of the Internet and its applications in recent years has resulted in creation of large-scale complex networks involving tens or hundreds of millions of nodes and links. 
Thus, it may be impossible or costly to obtain a complete picture of these large networks, and partially-observed data is often being studied for network characterization. The network resulting from such measurements may be thought of as a sample from a larger underlying network. However, sampling in a network context introduces various potential complications such as the bias of selecting the nodes in the procedure of  sampling \cite{Frank2011}.

Many network characterization metrics we are aware of can be expressed as averages of some target functions. For example, the density of infected computers with a virus in the Internet, the ratio of sport websites in the WWW networks, and the average number of copies of a file in a Peer-to-Peer network, can be computed by the average of the target function $f$: $f(x)=1$, if node $x$ has desired characteristic (respectively, is infected with that virus, is a sport website, has a copy of that file), and $f(x)=0$, otherwise. 

A measurement framework for computing the average of such functions can be achieved in two steps; (1) Sampling, and (2) Estimation. In many cases (e.g. the Internet), the global structure of a network is initially unknown and there is no sampling frame (i.e., a list of all nodes within a network, from which nodes can be directly sampled). Therefore, using classical sampling methods such as random sampling are either impossible or impractical, and link-tracing based sampling methods are the only feasible solutions to collect data from such networks. In these methods, one can see the neighbors of already sampled nodes and make a decision on which nodes to visit next. Snowball (which is similar to breadth-first search in computer science) \cite{Frank2011} and Random walk based methods \cite{Heckathorn1997} are among the most popular link-tracing based methods.

In the Estimation step, an estimator is used to approximate the network characteristics. An estimator is a function that accepts a summary of the sampled data as input and outputs an estimate of an unknown network characteristic. Specifically, the selection bias of a sampling method can be corrected by using the general idea of Hansen-Hurwitz estimator \cite{Hansen1943}, i.e. the measured value is weighted inversely proportional to the visiting probability of the corresponding node \cite{Volz2008}. 
Therefore, we need to compute (or approximate) the visiting probabilities of each sampled node. In particular, in undirected networks, when data from snowballs with more than one wave are available, it might be comparatively easy to compute the visiting probabilities of each sampled node based on the first few waves, but it is more complicated to find the required visiting probabilities for snowballs of many waves \cite{Frank2011, kurant2011towards}. For random walk based sampling methods, in a sufficiently long ergodic walk on an undirected network, we would expect that the probability that a node is visited is proportional to its degree \cite{Lovasz1993}.

Many of the networks around us such as WWW, Twitter, and Peer-to-Peer networks contain directed links, or links which do not have the same strength in each direction. However, despite recent efforts on characterizing undirected networks based on the sampled data \cite{Salehi2012, Gjoka2011a, Gjoka2011b, Ribeiro2010}, little attention has been made on the statistical properties of sampled directed networks \cite{Lu2011, Ribeiro2012}. The directed links pose a considerable challenge to the process of network measurement (i.e., unknown bias in the sampled nodes). In summary, computing (or approximating) the visiting probabilities of sampled nodes for directed networks still remains as an open problem. For example, the stationary distribution for random walks on a directed network is no longer determined by the degree sequences. We can only show that a random walk process on a strongly connected directed network has a unique stationary distribution (but, there is no analytical solution for it yet). 
However, most real-world directed networks are not strongly connected. Without this property, the limiting distribution of a random walk is sensitive to the starting distribution, and its analysis can be very complicated. A modification of random walk which overcomes this problem allows for a jump with small probability to a node at random from the whole network. That is called PageRank in the Internet search literature \cite{Brin1998}. It's clear that according to our assumptions about the absence of sampling frame, jumping to a random node from the whole network may not be possible. Therefore, PageRank can not be applied to our problem in its global definition.

In this paper, we propose a complete framework to measure nodal characteristics which can be defined by average target functions in an arbitrary directed network. In the sampling step, we introduced a novel link-tracing algorithm by utilizing the general idea of PageRank. We employ a modified version of PageRank, called personalized PageRank \cite{jeh2003}, using a specific set of nodes as a seed vector. Jumping is limited to this seed vector. Thus, personalized PageRank determines the importance of every node to the seed by using local information. In particular, the proposed method samples the underlying network by moving from a node to one of its neighbors through an outgoing link, based on the approximated probability of personalized PageRank. 

Since these probabilities represent an approximation of the exact visiting probabilities, we utilize the idea of approximate ratio importance sampling to propose a new estimator. 
The previous well-known work \cite{Volz2008, Salganik2004} on the estimators for correcting the bias, does not consider the effect of approximation of visiting probabilities which we address in this paper.
Importance sampling is one of the methods of Monte Carlo simulation \cite{Liu2001} in which instead of generating a sample from the target distribution (e.g., uniform distribution in our case), the estimator generates a sample from a different trial distribution (e.g., the distribution generated by the proposed sampling algorithm). We evaluate our framework over large synthetic and real datasets in terms of estimated average of a target function (here, we focus on the infection ratio and outdegree distribution) and its bias. Our results show that the proposed method performs well even in low sampling rates. 

\section{Related Work}
\label{sec:Related Work}
Link-tracing based sampling methods in the context of networks are somewhat distinct from classical sampling methods \cite{Frank2011}. Random walk (RW) \cite{Lovasz1993} is one of the most important and widely used sampling methods in different kind of network contexts such as uniformly sampling Web pages from the Internet \cite{Henzinger2000}, content density in Peer-to-Peer networks \cite{Stutzbach2008}, degree distributions of the Facebook social graph \cite{Gjoka2011b}, and collecting data from information diffusion networks \cite{Eslami2012a, Eslami2012b}. A classic RW, samples a graph by moving from a node, $x$, to a neighboring node, $y$, through an outgoing link, chosen uniformly at random from the neighbors of $x$. By this process links and nodes are sampled. 
In any given connected and non-bipartite undirected graph, the classic RW is biased towards node with higher degree. 

In general, the probability of selecting the next node determines the probability that nodes are visited in sampling procedure. Metropolis-Hastings technique \cite{Metropolis1953} can be used to modify the probabilities of the node selection in random walk in order to have the uniform stationary distribution for visiting each node. This technique is a general Markov Chain Monte Carlo (MCMC) method \cite{Chib1995} for sampling from probability distributions based on constructing a Markov Chain that has the desired distribution as its stationary distribution. This approach, known as  Metropolis-Hastings Random Walk (MHRW), has been applied to Peer-to-Peer networks \cite{Stutzbach2008}, Facebook \cite{Gjoka2011b}, and Twitter \cite{Salehi2011}. 

Alternatively, one can use the unmodified classic RW method to sample from a graph and correct the degree bias by re-weighting the sampled values. The Respondent-driven sampling (RDS) \cite{Heckathorn1997,Salganik2004} is an example of this approach. Actually, this is a framework in the field of social sciences to sample and infer in hard-to-reach populations such as injection drug users. In these populations, a sampling frame for the target population is not available. Sampling from OSNs or WWW network graph is analogous to the sampling of hidden population in the social sciences. In the context of graph sampling, the RDS method has been used for Facebook \cite{Gjoka2011b}, Twitter \cite{Salehi2011}, and Peer-to-Peer networks \cite{Stutzbach2008}. 
 
One of the main assumptions of the above mentioned studies is that the underlying network is undirected. Since most real-world networks (such as WWW, Twitter, and Peer-to-Peer networks) are directed, this assumption is not easily met in real life. However, little attention has been paid to directed networks by considering the effect of directed links. The authors in \cite{Hall2009} consider directed links in the Peer-to-Peer network while maintaining desirable statistical properties (i.e., uniformity) for the sampling procedure. The main idea of this RW based method is to avoid the calculation of each node's visiting probability, and instead to adjust the transition probabilities iteratively, converging to a state in which the sum of transition probabilities into each node equals $1$. The resulting transition matrix is said to be doubly stochastic, and induces uniform stationary distribution. 

In \cite{Lu2011}, it is shown that when the sample size is relatively small, the RDS method may generate relatively large biases and errors if the studied networks are directed, indicating that estimates from previous RDS studies should be interpreted and generalized with caution. In \cite{Ribeiro2012}, a random walk sampling algorithm with jumps, called Directed Unbiased Random Walk (DURW), is proposed that achieves asymptotically unbiased estimates of the outdegree distribution of a directed graph. The authors construct an undirected graph using the nodes that are selected by the random walker on the directed graph. The undirected graph is built to allow the walker to traverse known outgoing links backwards, and guarantees that the probability of sampling a node can be approximated, even though incoming links are not observed. However, this method is based on the assumption that nodes can be sampled uniformly at random from the original graph (i.e., there is some means of obtaining a list of nodes in the network) which is not always feasible. For example, while it is feasible for Wikipedia and Twitter, it is not feasible for the WWW graph.

In conclusion, while the effect of directed links have been studied, there is no efficient measurement framework for directed networks in the literature. Moreover, no analytical solution for the visiting probabilities of the nodes is available for a general directed network. 

\section{Preliminaries}
\label{sec:Preliminaries}
\subsection{Basic Notations and Definitions}
Let $G = (V, E)$ with $n=|V|$ and $m=|E|$ be the graph representing a complex network, where $V$ is the set of nodes, and $E=\{(i,j): i,j\in V, A_{ij}=1\}$ is the set of unweighted directed links between pairs of nodes. $A_{ij}$ indicates the presence ($=1$) or absence ($=0$) of  relation of interest (such as a hyper-link in WWW, friendship in a social network, and file transferring in a Peer-to-Peer network) from $i$ to $j$. 
The above $n \times n$ matrix $A$ is called an adjacency matrix. 
The indegree of a node $x$, $d_{in}(x)$, is the number of distinct ingoing links $(y_1 , x), . . . , (y_k , x)$ into $x$, and its outdegree, $d_{out}(x)$, is the number of distinct outgoing links $(x, y_1 ), . . . , (x, y_k)$ out of $x$. 

A link is called reciprocal if a connected pair of nodes have both an ingoing and an outgoing link between each other. 
The proportion of reciprocal links in a network called reciprocity, $r$ (i.e. $r = 1$ when the network is undirected, and $r = 0$ when the network is directed in a way such that there are no reciprocal links).

Let $S$ be a sample of nodes where $S \subset V$, and $G(S)$ is the induced subgraph of $G$ based on the sample $S$. 
Let $N(S)$ denote the neighborhood of $S$; 
\begin{equation} \label{NeighborhoodEqu}
N(S) = \{w \in V \-- S: \exists v \in S \hspace{2mm} s.t. \hspace{2mm} (v,w) \in E\}
\end{equation}
Therefore, $N(S)$ is the set of nodes that we know exist due to other outgoing links but have not yet visited. 

Let $L$ be a set of real-valued labels. For instance, a label can be the degree of a node. We assume that a label $l_x \in L$ is assigned to each node $x \in V$ by a target function $f:V \rightarrow L$, i.e. $f \subseteq \{(x,l_x): x\in V, l_x \in L\}$. We can express many network characterization metrics as averages of some target functions over $V$. Given a target function $f$, the average of $f$ is: $avg(f)={\sum_{x \in V} f(x)/ |V|}$.

Our main goal in this paper is to propose a measurement framework to compute $avg(f)$ (based on the sampled nodes $S$) for a general target function $f$ (that defines a nodal characteristic) in an arbitrary directed network $G$. Our only assumption on $G$ is that it is weakly connected, i.e. every node can be reached from every other node by traversing the links without considering the direction of the links. 

\subsection{Personalized PageRank Vectors}
PageRank was first introduced by Brin and Page \cite{Brin1998} for search algorithms in the WWW network. However, it can be defined for any graph as a stationary distribution of a certain random walk over that graph. At each step, with probability $(1-\alpha)$, the random walk follows a randomly outgoing link of a node, and with probability $\alpha$ the random walk makes a jump to a new node chosen uniformly among all nodes in the network. The jumping constant $\alpha$ ($0 < \alpha \leq1$), controls the diffusion rate. With smaller $\alpha$, the random walks spread further from the initial nodes before performing a random jump. Traditionally, the value of $\alpha$ is chosen to be $0.15$.

Let $A$ be the adjacency matrix of a graph $G$. The PageRank vector $\mathbf{p(\mathbf{s})}$ of this graph is a visiting probability distribution on the nodes of $G$ that can be defined as the solution of the following equation:
\begin{equation}
\label{PageRankEqu}
{\mathbf{p(\mathbf{s})}} = \alpha{\mathbf{s}}+(1-\alpha){\mathbf{p} D_{A}^{-1}A}
\end{equation}
where $\mathbf{s}$ is a seed vector that includes the initial distribution for starting nodes. 
We use the notation $D_{A}^{-1}$ to indicate the diagonal matrix with the inverse outdegree for each node. Globally, uniform $\mathbf{s}=\mathbf{1}/n$ is considered for computing PageRank. However, an arbitrary distribution ${\mathbf{s}}$ can lead to creation of personalized PageRank \cite{Haveliwala2003}. Specifically, the random walk in the personalized PageRank only jump to a few nodes of personal interest. It can be shown that for any ${\mathbf{s}}$ and $\alpha$, there is an unique vector $\mathbf{p(\mathbf{s})}$ satisfying Equ.(\ref{PageRankEqu}) \cite{jeh2003}. 

There are various algorithms for computing global PageRank and personalized PageRank \cite {jeh2003, berkhin2006, gleich2006}. As a standard PageRank algorithm, we can efficiently compute the solution of Equ.(\ref{PageRankEqu}) by applying the power method \cite{Golub1996}. 
Given some initial distribution  $\mathbf{s}$, the power method is defined as following iteration:
\begin{center}
${{\mathbf{p}^{(k+1)}} = (1-\alpha)  {\mathbf{p}^{(k)}} D_{A}^{-1} A}$
\\
${{\mathbf{p}^{(k+1)}} = {\mathbf{p}^{(k)}} + (1- {\| {\mathbf{p}}^{(k+1)} \|_{1}})\mathbf{s}}$
\end{center}
where $ {\| .\|_{1}}$ denotes the $1$-norm. As an algorithm, the power method continues this iteration until  ${\|{\mathbf{p}^{(k+1)}} - {\mathbf{p}^{(k+1)}}\|_{1}} < \delta$  for a user-provided stopping tolerance $\delta$.   	 
Based on the idea of the power method, Gleich and Polito \cite{gleich2006} present an efficient algorithm, called boundary-restricted personalized PageRank algorithm (BRPPR), to compute an approximation of the personalized PageRank vector $\mathbf{p(\mathbf{s}})$ by examining only a small fraction of the input web graph near the starting vector $\mathbf{s}$. Specifically, this algorithm iteratively divide the web pages into an active and inactive set. At each iteration, the set of active web pages is expanded to include more web pages that are likely to have a high personalized PageRank value. 
Specifically, it expands pages until the total rank on the frontier set (the set of pages that we know exist due to other outgoing links but have not yet visited) of pages is less than an expansion tolerance. 
Therefore, only the set of active pages and their outgoing link information are actually involved in the computation. Moreover, a stopping tolerance threshold is considered in order to determine the termination of the approximation algorithm. The analysis of BRPPR leads to the following theorem (theorem 3.3 and remark 3.4 of \cite{gleich2006}):
\begin{mytheorem}
\label{Theorem:ApproximatePR}
The BRPPR algorithm with expansion tolerance $\kappa$ and stopping tolerance $\delta$ yields an approximate PageRank vector $\hat{\mathbf{p}}$, where $ {\| {\mathbf{p}} - \hat{\mathbf{p}}\|_{1}} \leq {{2(1-\alpha)}\over{\alpha}}. \kappa +{{2-\alpha}\over{\alpha^2}}.\delta $\\
\end{mytheorem}
where ${\| {\mathbf{p}} - \hat{\mathbf{p}}\|_{1}}$ gives the Manhattan distance between two distribution vectors of $\mathbf{p}$ and $\hat{\mathbf{p}}$. Although, the authors in \cite{gleich2006} address the problem of calculating the PageRank score of a Internet webpage in the field of search engines, we utilize their general idea here in the context of network sampling. 

\subsection{Importance Sampling}
\label{sec:Approximate Importance Sampling}
As mentioned, we are interested in measurement of quantities that can be written as averages of some target functions $f$ over a finite set $V$ (i.e., the set of all nodes).
Let $\pi:V\rightarrow [0, \infty)$ be the target distribution on a set $V$. In our problem, the desired $\pi$ is the uniform distribution on $V$, then all nodes $x$ are equally likely to be selected, i.e. $\pi(x)=1/|V|$. Let $\hat{\pi}: V\rightarrow [0, \infty)$ be a target measure that is unnormalized form of $\pi$, i.e. $\hat{\pi}(x)={\pi(x)}.{Z_{\hat{\pi}}}$ where $Z_{\hat{\pi}}>0$ is the normalization constant of ${\hat{\pi}}$. We say that measure ${\hat{\pi}}$ induces a corresponding probability distribution on $V$.
Therefore, ${\hat{\pi}}$ is a relative weight, which represents the probability of $x$ to be chosen in the distribution $\pi$. In our problem, we have ${\hat{\pi}}=1$ (for all $x \in V$) and $Z_{\hat{\pi}} = |V|$. 

The average of a target function relative to $\hat{\pi}$ is essentially a sum where the target measure is the probability distribution $\pi$. Hence, we can write
\begin{equation}
avg(f)=avg_{\hat{\pi}}(f)=sum_{{\pi}}(f)={\sum_{x\in V} f(x).{{\pi}(x)}}
\end{equation}

For example, in our previous example of infection in a population, the infection ratio is an average of the target function relative to an uniform target measure. Alternatively, it is a sum of the same function relative to the uniform distribution on $V$ (i.e., $\pi(x) = 1/|V|$ for all $x$).

The naive estimator for $avg_{\hat{\pi}}(f)$ works as follows \cite{Liu2001}: (1) Generate a random sample $X$ from the distribution $\pi$ induced by $\hat{\pi}$, (2) Compute the function of $f$ for sampled $X$ (i.e., $f(X)$).
It is easy to check that this estimator is unbiased. However, since we assume that there is no sampling frame in our problem, uniform sampling from the distribution $\pi$ may be hard or costly. Consequently, this simple estimator is inefficient. To solve this problem, we can use \textit{Importance Sampling (IS)}, as one of the methods of monte carlo simulation \cite{Liu2001}. In the following, we state some theorems about the general idea of importance sampling \cite{Gurevich2011}. Although, the authors in \cite{Gurevich2011} address the problem of externally measuring aggregate functions over documents indexed by search engines, we use them here in the context of network sampling. 

\subsubsection{Importance Sampling (IS)}
The basic idea of importance sampling is: instead of generating a sample $Y$ from the target distribution $\pi$, the estimator generates a sample $X$ from a different trial distribution $p$ on $V$ to directly estimate statistical sums relative to the target measure $\hat{\pi}$. Specifically, given a target function $f$, importance sampling estimator estimates $sum_{\hat{\pi}}(f)={\sum_{x\in V} f(x).{\hat{\pi}(x)}}$. 
The trial distribution $p$ can be any distribution, as long as $supp(\pi) \subseteq supp(p)$ (here, $supp(p)=\{x \in V | p(x)>0\}$ and $supp(\pi)$ is defined similarly). In particular, we can choose $p$ to be a distribution that is easy to sample from. By considering $\hat{p}$ as unnormalized form of $p$, the importance sampling estimator is then defined as follows:
\begin{equation}
E_{IS}(X)=f(X).{{\hat{\pi}(X)}\over{\hat{p}(X)}}=f(X).w(X), \hspace{5mm} X\in V
\end{equation}
where $w(X)$ is called the importance weight. Function $f$ is computed by samples of distribution $p$. 
It can be shown that $E_{IS}(X)$ is an unbiased estimator for $sum_{\hat{\pi}}(f)$. 
Implementation of an importance sampling estimator requires: (1) Ability to sample efficiently from the trial distribution $p$; and (2) Ability to compute the importance weight $w(x)$ (or its estimator) and the function value $f(x)$, for any given node $x \in V$. There is no need to know the normalization constant $Z_{\hat{\pi}}$ or to be able to sample from $\pi$. This basic importance sampling estimator is only suitable for estimating sums. We next extend it for estimating averages.

\textbf{Importance Sampling for average:}
Recall that the estimator for averages estimates $avg_{\hat{\pi}}(f) = sum_{\pi}(f)$. If we are able to compute $\pi(x)$ for each $x \in V$, we can use the importance sampling estimator described above to estimate average of a function relative to $\hat{\pi}$. However, in many cases we can only compute  $\hat{\pi}(x)$ and not $\pi(x)$. For example, in our case, $\pi(x) = 1/|V|$, where $|V|$ is the number of nodes, that is typically unknown.
Ratio importance sampling is a technique for estimating averages relative to any target measure $\hat{\pi}$, even when $\pi(x)$ is unknown. 

The ratio importance sampling estimator computes two sum estimators: $M_1$ of $sum_{\hat{\pi}}(f)$, and $M_2$ of $sum_{\hat{\pi}}(w)$, and then outputs $M_1/M_2$. 
However, there is one problem: 
the expectation of a ratio is not the ratio of the expectations, i.e., $\mathds{E}(M_1/M_2) \neq \mathds{E}(M_1)/\mathds{E}(M_2)$. 
To solve this problem, ratio importance sampling utilize the following idea: if we replace $\mathds{E}(M_1)$ and $\mathds{E}(M_2)$ by averages of multiple independent instances of the estimator of $\mathds{E}(M_1)$ and $\mathds{E}(M_2)$, the difference between $\mathds{E}(M_1/M_2)$ and $\mathds{E}(M_1)/\mathds{E}(M_2)$ diminishes to $0$. Therefore, the ratio importance sampling estimator for $avg_{\hat{\pi}}(f)$ is then defined as follows:
\begin{equation}
E_{RIS}(X_1,X_2,...,X_n) = {{{1 \over n} {\sum\limits_{i=1}^n E_{IS}(X_i)}} \over {{1 \over n} {\sum\limits_{i=1}^n w(X_i)}}}
\end{equation}
where $X_1,X_2,...,X_n$ are $n$ independent samples from the trial distribution $p$. 
\subsubsection{Approximate Importance Sampling (AIS)}
Unfortunately, in many cases we are unable to accurately compute the importance weight function $w$. This problem arises, e.g. in our case, due to complexity of analysing a RW based sampling method on a directed network. In the following, we address the effect of the approximate importance weights on importance sampling.
Approximate importance sampling method employ an \lq\lq{approximate importance weight function}\rq\rq   $u(x)$ rather than the exact one $w(x)$. This estimator can be defined as:
\begin{equation}
E_{AIS}(X)=f(X).u(X) 
\end{equation}
where $X$ is distributed according to the trial distribution $\hat{p}$. The estimation generated by approximate importance sampling is close to the true value as long as the importance weight function $w(x)$ and the approximate importance weight function $u(x)$ are similar. To analyse the bias of this estimator, we consider the following theorem.
\begin{mytheorem}
\label{Theorem:biasOfEST_AIS}
(\cite{Gurevich2011}) Suppose $supp(u) \subseteq supp(w)$\\
${\mathds{E}(E_{AIS}(X))}=sum_{\hat{\pi}}(f(Y)).\mathds{E}({u(Y) \over {w(Y)}})+Z_{\hat{\pi}}.cov(f(Y),{u(Y) \over w(Y)})$\\
\end{mytheorem}
where $X \sim {\hat{p}}$, $Y \sim \pi$, and $Z_{\hat{\pi}}$ is the normalization constant. Therefore, there are two types of bias in this estimator: (1) multiplicative bias, depending on the expectation of $u/w$ relative to $\pi$. It may have a significant effect on the estimator's bias, and thus must be removed. (2) additive bias, depending on the correlation between $f$ and $u/w$ and on the normalization constant $Z_{\hat{\pi}}$. It is typically less significant, as in many practical situations $f$ and $u/w$ are uncorrelated (e.g., when $f$ is a constant function, as is the case with infection ratio example). 

To remove the multiplicative bias, we assume that it is possible to estimate the multiplicative bias, and divide $E_{AIS}(X)$ by this estimate. For an instance $x \in V$, the ratio $u(x)/w(x)$ is called the weight skew at $x$. The multiplicative bias factor is the expected weight skew relative to the target distribution $\pi$. 
To remove this bias, we need to estimate the expected weight skew. Let $E_{WSE}(X)$ be an unbiased weight skew estimator for $\mathds{E}({{u(Y)} \over {w(Y)}})$. 
It follows from Theorem \ref{Theorem:biasOfEST_AIS} that:
\begin{equation}
{{\mathds{E}(E_{AIS}(X))} \over {\mathds{E}(E_{WSE}(X))}}=sum_{\hat{\pi}}(f(Y))+{Z_{\hat{\pi}}.{cov(f(Y),{u(Y) \over w(Y)}} \over {\mathds{E}({u(Y) \over {w(Y)}})}}
\end{equation}
Thus, the ratio of the expectations of the two estimators, $EST_{AIS}(X)$ and $EST_{WSE}(X)$, gives us the desired result $sum_{\hat{\pi}}(f)$, modulo an additive bias factor that depends on the correlation between $f$ and $u/w$. The additive bias can be considered $0$ since $f$ is often a constant function. Ignoring this additive bias, it would seem that the ratio ${EST_{AIS}(X)} / {EST_{WSE}(X)}$ is a good estimator for $sum_{\hat{\pi}}(f)$. 
Although the expectation of a ratio is not the ratio of the expectations, one applies the ratio estimation technique similarly to what we did in the ratio importance sampling. 
Therefore, the approximate ratio importance sampling estimator for $sum_{\hat{\pi}}(f)$ can be defined as follows:
\begin{equation}
E_{ARIS}(X_1,X_2,...,X_n) = {{{1 \over n} {\sum\limits_{i=1}^n E_{AIS}(X_i)}} \over {{1 \over n} {\sum\limits_{i=1}^n E_{WSE}(X_i)}}}
\end{equation}
where $X_1,...,X_n$ are $n$ independent samples from the trial distribution $\hat{p}$. 

\textbf{Approximate importance sampling for averages:}
Here, we employ $E_{ARIS}$ to design an importance sampling estimator for averages. Recall that the estimator for averages estimates $avg_{\hat{\pi}}(f) = sum_{\pi}(f)$. Thus, we need to compute the importance weight function $w_{avg}(x)={\pi}(x)/{\hat{p}}(x)$.
However, $\pi(x)$ cannot be calculated exactly. For solving this problem, we use an approximate weight function $u(x) \approx w_{avg}(x)$ instead of $w_{avg}(x)$, and then fixing the bias using approximate ratio importance sampling estimator (i.e., $E_{ARIS}$) described above. To this end, we need to come up with the weight skew estimator whose expectation equals $\mathds{E}(u(Y)/w_{avg}(Y))$, where $Y$ is distributed according to $\pi$. It can be shown that with a constant function $f$, the approximate weight $u(X)$ itself is an unbiased estimator for $\mathds{E}(u(Y)/w_{avg(Y)})$ \cite{Gurevich2011} . We thus set $E_{WSE}(X) = u(X)$. Therefore, approximate importance sampling for averages, $E_{AVG}$, is then defined as follows:
\begin{equation}
\label{equ:AIS_avg}
E_{AVG} = {{{1 \over n} {\sum\limits_{i=1}^n f(X_i).u(X_i)}} \over {{1 \over n} {\sum\limits_{i=1}^n u(X_i)}}}
\end{equation}
where $X_1,...,X_n$ are $n$ independent samples from the trial distribution $\hat{p}$. The following theorem shows the bias and the variance of this estimator.
\begin{mytheorem}
\label{Theorem:biasOfE_AVG}
(\cite{Gurevich2011}) The bias and variance of $E_{AVG} $ (equ. \ref{equ:AIS_avg}) are defined as
\\ 
$Bias(E_{AVG})=E_{AVG}(X_1,X_2,...,X_n) - avg_{\hat{\pi}}(f)={Z_{\hat{\pi}}.{cov(f(Y),{u(Y) \over w_{avg}(Y)})} \over {\mathds{E}({u(Y) \over {w_{avg}(Y)}})}} + {O({1 \over n})}$
\\
$var(E_{AVG}(X_1,X_2,...,X_n))={O({1 \over n})}$
\end{mytheorem}

\section{Proposed Measurement Framework}
\label{sec:ProposedMeasurmentProcess}
The pseudo code of proposed framework, called Directed-Network-Measurement (DNM), is shown in Algorithm \ref{DNM-algorithm}. 
Our main goal is to approximate the probability of each visited node in sampling procedure, and correct the bias of sampling by using the idea of approximate importance sampling. 
The proposed framework is designed through two steps; sampling and estimation. The sampling step initially starts with an empty sample set $S$, and adds a random node $y$ to it as a seed. Subsequent nodes of the sample are chosen based on the following procedure. In each iteration, the set of neighborhood of already sampled nodes, i.e. $N(S)$, and its corresponding active link matrix, $L$, is generated. The matrix $L$ is the adjacency matrix for all outgoing links from nodes in $S$, but may reference nodes that are not in $S$. We then compute one PageRank iteration on $L$ from a vector $\hat{\mathbf{p}}$ to a vector $\tilde{\mathbf{p}}$. Next, we add nodes (that are likely to have a high personalized PageRank value) to $S$ until the total probability on the remainder of the neighborhood is less than an expansion ratio of $\kappa$. 

As we mentioned earlier, the essential property of a sampling method that makes it appropriate for network inference is that its visiting probabilities are known or estimable for the sampled nodes. In the estimation step, we use computed PageRank value of each sampled node (i.e. $\hat{\mathbf{p}}$), as an approximation of the exact visiting probability. Then, we utilize the idea of approximate importance sampling to design a new estimator and study its performance for a general target function. Specifically, we consider the obtained $\hat{\mathbf{p}}(x)$ as an approximate weight function $u(x)$. 

According to the general ratio importance sampling estimator (refer to Equ.\ref{equ:AIS_avg}), we propose the following estimator for computing the average of a general target function $f$ in a directed network.
\begin{equation}
\label{est:dir}
E_{DIR} = {{{1 \over n} {\sum\limits_{i=1}^n f(X_i).\hat{\mathbf{p}}(X_i)}} \over {{1 \over n} {\sum\limits_{i=1}^n \hat{\mathbf{p}}(X_i)}}}
\end{equation}
where $X_1, X_2, ..., X_n$ are $n$ samples extracted by sampling procedure.  

\begin{pseudocode}[ruled]{DNM}{y, \alpha, \kappa, \delta} \label{DNM-algorithm}
    \mbox{$S = \{y\}$; $L$ = matrix of $y$ to its outgoing links}\\
    \mbox{SUM = $0$}\\
    \mbox{WSE = $0$}\\
   	\mbox{$\hat{\mathbf{p}}=0; \hat{\mathbf{p}}(y) = 1$}\\
   	   \mbox{\% Sampling step} \\
   	\REPEAT \DO
   	\BEGIN  
   	\mbox{$\tilde{\mathbf{p}} = (1-\alpha) \hat{\mathbf{p}} D_{L}^{-1} L$}\\
	\mbox{$\tilde{\mathbf{p}} = \tilde{\mathbf{p}} + (1- {\| \tilde{\mathbf{p}} \|_{1}}$)}\\
	 \WHILE {\tilde{\mathbf{p}}(N(S)) > \kappa} \DO 
	 \BEGIN	
	 \mbox{Find $x \in N(S)$ with max value in $\tilde{\mathbf{p}}$}\\
	 \mbox{Add $x$ to $S$}\\
 	 \mbox{Remove $x$ from $N(S)$}\\
	 \mbox{Update $\tilde{\mathbf{p}}(N(S))$}\\
	 \END\\
 	 \mbox{Update $D_L$ and $L$}\\
     \mbox{$\omega = \|\tilde{\mathbf{p}} - \hat{\mathbf{p}}\|_{1}$}\\ 	 
 	  \mbox{$\hat{\mathbf{p}} = \tilde{\mathbf{p}}$}\\
 	  \END\\
 	 \UNTIL{\omega < \delta}\\
 	 \mbox{\% Estimation step} \\
 	 \FOR {x \in S} \DO
 	 \BEGIN
 	 \mbox{SUM = SUM + $f(x).\hat{\mathbf{p}}(x)$}\\
	 \mbox{WSE = WSE + $\hat{\mathbf{p}}(x)$}\\
 	 \END\\
	 \RETURN{\mbox{SUM/WSE}}
\end{pseudocode}

The following theorem shows the bias and variance of the proposed estimator.
\begin{mytheorem}
\label{Theorem:estDNM} 
 The bias and variance of $E_{DIR} $ (equ. \ref{est:dir}) are defined as\\
$Bias(E_{DIR})=E_{DIR}(X_1,X_2,...,X_n) - avg_{\hat{\pi}}(f)={{|V|}.{cov(f(Y),{{\hat{\mathbf{p}}} / {\mathbf{p}}})} \over {\mathds{E}({{\hat{\mathbf{p}}} / {{\mathbf{p}}}})}} + {O({1 \over n})}$
\\
$var(E_{DIR}(X_1,X_2,...,X_n))={O({1 \over n})}$\\
\end{mytheorem}
where $Y \sim \pi=1/|V|$, and $Z_{\hat{\pi}}$ is the normalization constant.

\textbf{proof:}
By substituting $u=1/|V|$, $w_{avg}={\mathbf{p}}$, and $Z_{\hat{\pi}}=|V|$ into Theorem \ref{Theorem:biasOfE_AVG}, the theorem is proved.

We conclude from this theorem that if we use sufficiently large number of samples, then we are likely to obtain an estimate of $avg_{\hat{\pi}}(f)$, which has only additive bias that depends on the correlation between $f$ and ${{\hat{\mathbf{p}}} / {{\mathbf{p}}}}$.

\section{Experimental Evaluation}
\label{sec:expEval}

\subsection{Datasets}
\label{subsec:Datasets}

\textbf{Synthetic Networks:} We utilize three kinds of random models for generating directed networks:

Directed Erdos-Renyi (DER) networks \cite{Erdos59}: The model starts with $|V|$ nodes, which link to each other with probability $p_0 \in (0, 1)$. The reciprocity of these network equals to $1$. In order to decrease the reciprocity, reciprocal links in the base network are randomly chosen and converted to irreciprocal links. We generate networks by using this model with $|V|=2000$, $p_0=0.1$, and $r=0.6$. 

Directed Watts-Strogatz (DWS) networks \cite{Watts98}: The model starts from a completely regular network with identical degree ($k$-nearest neighbors' connection) and clockwise links. Each link will be rewired with two randomly selected nodes with probability $p_0 \in (0, 1)$. In simulation, we generate networks with $|V|=2000$ nodes by setting  $k=20$ and $p_0=0.1$.  

Directed Scale Free (DSF) \cite{Pennock2002}: 
Network starts with $m_0$ nodes, which link to each other with probability $p_0$ (as in an Erdos-Renyi random graph). At each time step, one node and $m$ links are added to the network. The endpoints of the links are randomly selected among all nodes according to the following probability: $p_1(x) = \beta_1 {{d_{in}(x)}\over{|E|}}+ \beta_2 {{d_{out}(x)}\over{|E|}}+\beta_3 {{1}\over{|V|}}$, where $\beta_1 + \beta_2 + \beta_3 = 1$. We generate networks with $|V|=1000$ and average degree of nodes $=50$. We leave the number of links, $|E|$, unconstrained. For attaching probability, we set $\beta_1=0.7, \beta_2=0.2, \beta_3 = 0.1$. Moreover, we set $m_0=2$ for generating initial Erdos-Renyi graph and consider $p_0=1$ to have them fully connected.

\textbf{Real-world Networks:} Moreover, we considered some real-world directed network: 
 
An Online Social Network (OSN): This is a Facebook-like Social Network \cite{Opsahl2009} that originate from an online community for students at University of California, Irvine. The dataset includes the 1,899 users that sent or received at least one message. There are 20,296 directed links among these users. 

A Peer-to-Peer Network (P2P): We use a snapshot of the Gnutella Peer-to-Peer file sharing network collected in August 2002 \cite{Leskovec2007a}. Nodes and links represent hosts and connections between them in this network, respectively. The dataset includes the 10,876 hosts and 39,994 directed links.

The Internet Autonomous System (AS): The network of routers comprising the Internet can be organized into sub-networks called Autonomous Systems (AS). Each AS exchanges traffic flows with some neighbors. We use a CAIDA AS graph collected in February 2004 \cite{Leskovec2005}. The dataset contains 16,493 nodes and 66,744 links.

\subsection{Target Function}
\label{subsec:targetFuncion}

The fast growth of Internet and other communication networks makes them a suitable target for malicious activities. An infection spreads through the links of such networks constructed by computers and their communication channels. The process by which malicious objects such as worms, trojan horses, and computer viruses travel through computer networks is analogous to the process of spreading epidemics through a population. In this paper, we study the infection ratio in spreading of a disease in a population as a network characteristic. It is an example of family of binary properties of a network where each node is tagged by label $0$ or $1$. In particular, infection ratio can be considered as the average of a target function $f$: $f(x)=1$, if individual $x$ is infected, and $f(x)=0$, otherwise.

To spread disease on the test datasets, we use the susceptible-infectious-recovery (SIR) epidemic spreading model \cite{Moreno2002}. This is an epidemiological model widely used to simulate the spreading of epidemics, i.e. number of people infected with a contagious disease, in a population as a function of time. At every time step, the model assumes a transition rate $\theta_1$ for a susceptible person to become infected, if an infected neighbor is present, and a rate $\theta_2$ for an infected person to become recovered or die. The recovered person will never be infected again. In the simulations, we use $\theta_1=0.2$ and $\theta_2=0.05$. We run the simulation until desired ratio of infected nodes obtained.

\subsection{Experimental Setups}
\label{subsec:Experimental Setups}

We evaluate the performance of the DNM framework in various situations and in terms of different aspects. We consider two metrics; $1)$ estimated infection ratio, and $2)$ bias (which is defined by the absolute difference between estimated and true infection ratio). 
For each sampling rates in the range of $0.01\%$ to $40\%$, we repeated the simulations 100 times by selecting random nodes as seed, and finally, averaged the obtained values. We set jumping factor $\alpha=0.15$ and stopping tolerance $\delta=10^{-7}$ as the inputs of the proposed framework. 

Moreover, we set expansion tolerance $\kappa$ based on the desired sampling rates. Since, the parameter $\kappa$ controls the accuracy of the approximation of PageRank values (i.e., the visiting probabilities of the sampled nodes), different values of $\kappa$ provides different  sampling rates. Specifically, the less $\kappa$ leads to more samples, and consequently the more accuracy is obtained. We study the obtained $\kappa$ per various sampling rates for all test networks. Our result is demonstrated in Figure $1$. 
As we can see, the desired sampling rate depends on the structure of underlying network. According to Theorem \ref{Theorem:ApproximatePR}, the maximum error of the approximating visiting probabilities is in the range of $(0,6.8)$. This means that the proposed sampling method computes the visiting probabilities of the sampled nodes with reasonable accuracy even in low and medium sampling rates.

\begin{figure}[tp]
\begin{center}
\includegraphics[scale=0.29]{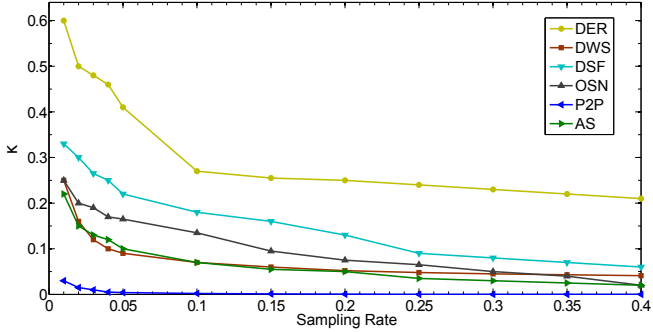}
\end{center}
\caption {The relation between expansion tolerance ($\kappa$) of the proposed sampling method and sampling rate for all test networks.}
\label{fig:spk}
\end{figure}

\subsection{Evaluation Results}
\label{subsec:Evaluation Results}

\subsubsection{Main Results}
\label{results}

\begin{figure}[tp]
  \begin{center}
    \subfigure[]{\label{AE-label}\includegraphics[scale=0.29]{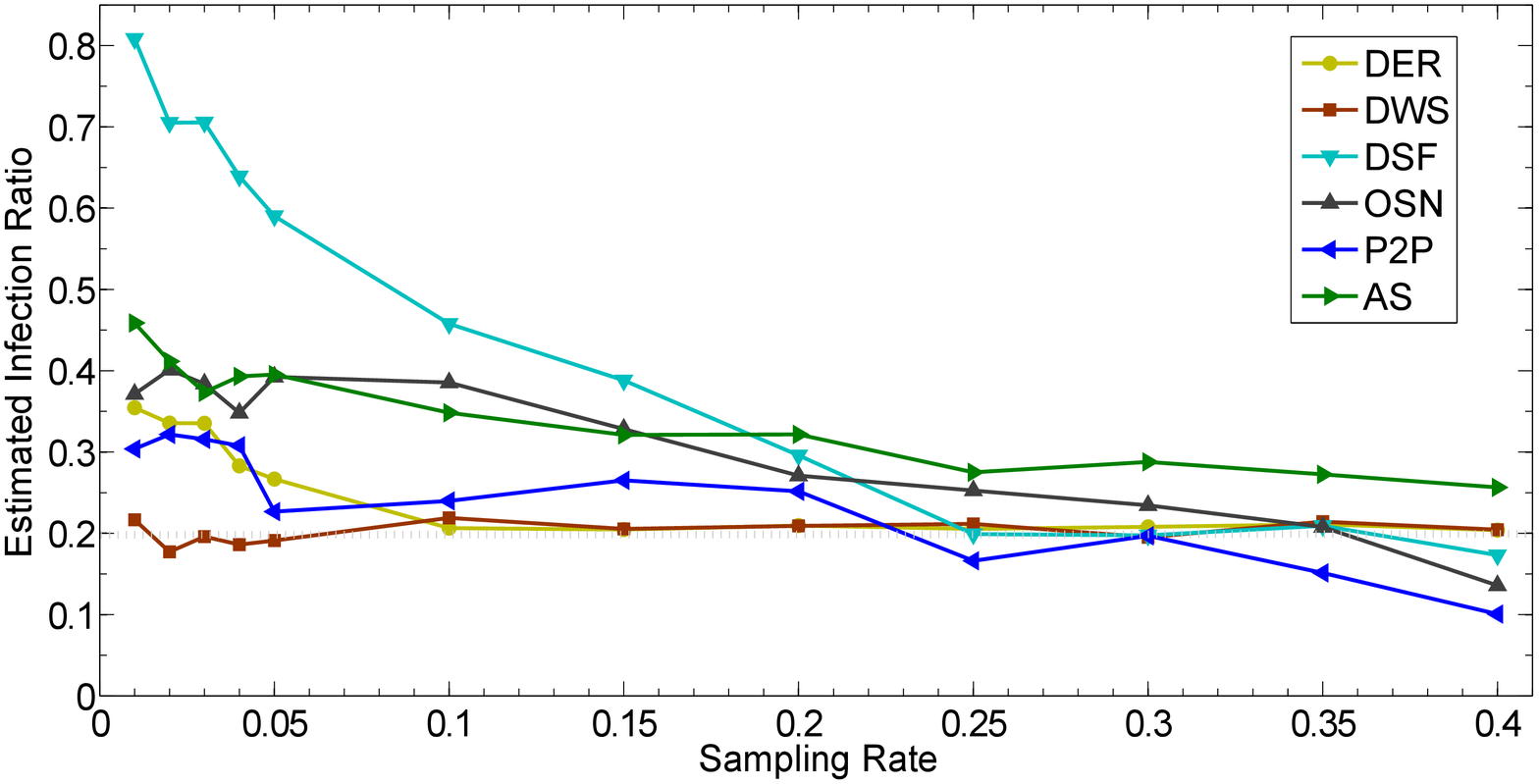}}
    \\
    \subfigure[]{\label{Bias-label}\includegraphics[scale=0.29]{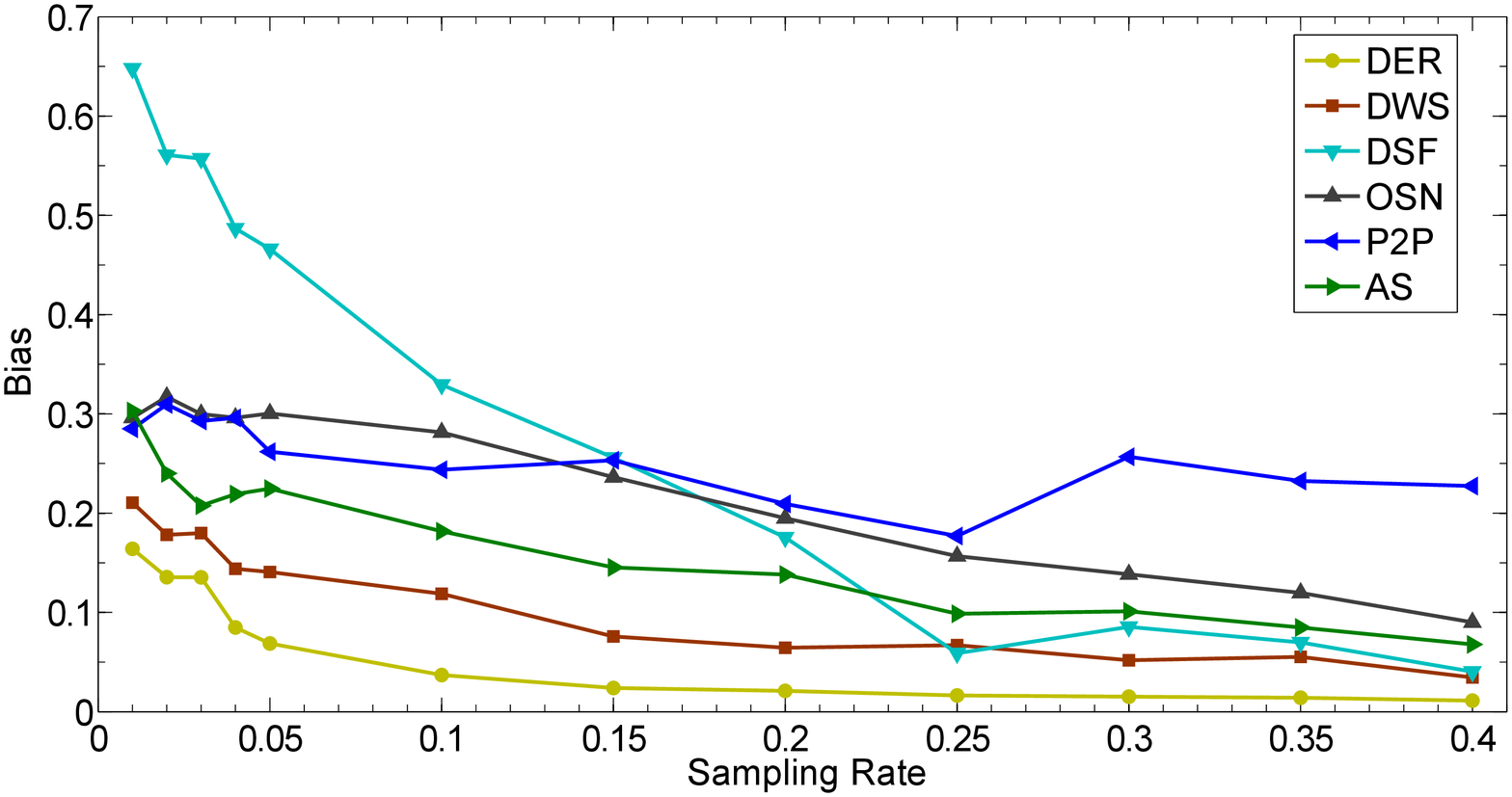}}
\end{center}
  \caption{The performance of the proposed framework in terms of (a) Estimated Infection Ratio (the true value is $20\%$) and (b) Bias.} 
\label{AE-main}
\end{figure}

The results of our study on the performance of proposed framework is shown in Figure \ref{AE-main}. As we can see, in general, the error in terms of estimated infection ratio and bias decreases significantly by increasing the sampled nodes. In particular, we reach desirable result in medium sample rates (estimated infection ratio and bias converges to $20\%$ and $0$, respectively). 
However, we found different behaviour in DWS networks; the estimated infection ratio is close to true value in all sampling rates (i.e. it's independent of number of samples). Moreover, the bias decreases with increasing the sampling rate.

\subsubsection{Convergence analysis}

Deriving valid population estimates from sampled nodes is based on the assumption that the samples are derived from the equilibrium distribution, which is true asymptotically. In this section, we study the convergence of the DNM estimates to equilibrium distribution during the data collection step.
 
In general, the starting nodes (the seeds) are not selected from a sampling frame, but instead are ad-hoc samples. One way to reduce the dependence of final estimates to seed nodes is to use a burn-in period by discarding large numbers of initial sampled nodes before analyzing the collected data. 
Given the high cost in terms of time and effort of collecting data, this may not be a desirable approach. Moreover, the authors in \cite{Gile2010} demonstrate that in a without-replacement sampling setting, this approach can even introduce more bias. In fact, the only real way to apply a burn-in to ensure accuracy would be to repeat the burn-in after every sampled node.

An alternative approach would be to estimate the relative visiting probabilities of all sampled nodes, conditioned on the composition of the seeds, and compute the estimates based on those probabilities. We follow this approach in the proposed framework. In particular, since jumping in personalized PageRank is limited to the seed vector, we approximate the importance of every node to the seed, i.e. the visiting probability, which is used in the Estimation step to correct the bias to the seeds. 

\begin{figure}[htp]
\begin{center}
\includegraphics[scale=0.29]{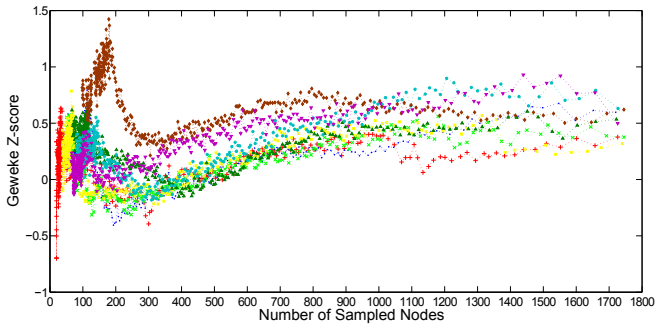}
\end{center}
\caption {The Geweke convergence diagnostic of DNM for infection ratio in the OSN dataset. The lines show the Geweke $Z$-score for some runs with different random nodes as seed.}
\label{fig:geweke1}
\end{figure}

To monitor the convergence of the proposed framework, we use a standard diagnostic test developed within the MCMC literature, namely Geweke \cite{Geweke1992}. This test was applied for the first time in the context of network sampling in \cite{Gjoka2011b}. The Geweke diagnostic detects the convergence of a single Markov chain by comparing the location of the sampled parameter on two different time intervals of the chain. The test is a standard Z-score with the standard errors adjusted for autocorrelation. We used this diagnostic in different runs of the proposed framework (generated by selecting some random nodes as seed) for the OSN dataset, and compares the difference between the first $10\%$ and the last $50\%$ of the samples. Figure \ref{fig:geweke1} presents the results of the Geweke diagnostic for the infection ratio as a network characteristic. We observe that after sampling approximately $200$ nodes, the $Z$-scores are strictly between $[-1, 1]$. This indicates that the proposed framework has achieved a good mixing with our initial selection of the random seeds.

\subsubsection{The effect of network reciprocity}
\label{dir}

\begin{figure}[htp]
  \begin{center}
  \includegraphics[scale=0.29]{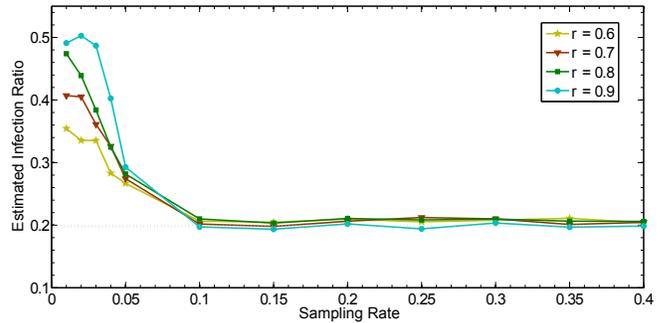} 
 \end{center}
  \caption{The effect of reciprocity ($r$) on the performance of DNM in underlying networks generated by DER model. True value of infection ratio (= $0.2$) shown in the grey dotted line.}  
  \label{Dir-whole}
\end{figure}

Here, we study the performance of the proposed framework in networks with different levels of reciprocity $r$ which is an important parameter in characterization of directed networks. To this end, we generate some underlying directed networks with $r \in \{0.6, 0.7, 0.8, 0.9\}$ by DER model. As we can see the results in Figure \ref{Dir-whole}, in lower sampling rates, the amount of error in estimation of infection ratio decreases with decreasing reciprocity (we observe the same pattern in terms of bias). As a general conclusion, if we have to sample a few nodes from a network, lower proportion of reciprocal links in that network leads to more accurate outputs.

\subsubsection{Sensitivity to true infection ratio}
\label{inf}

\begin{figure*}[tp]
  \begin{center}
    \subfigure[DER]{\label{DER-AE}\includegraphics[scale=0.27]{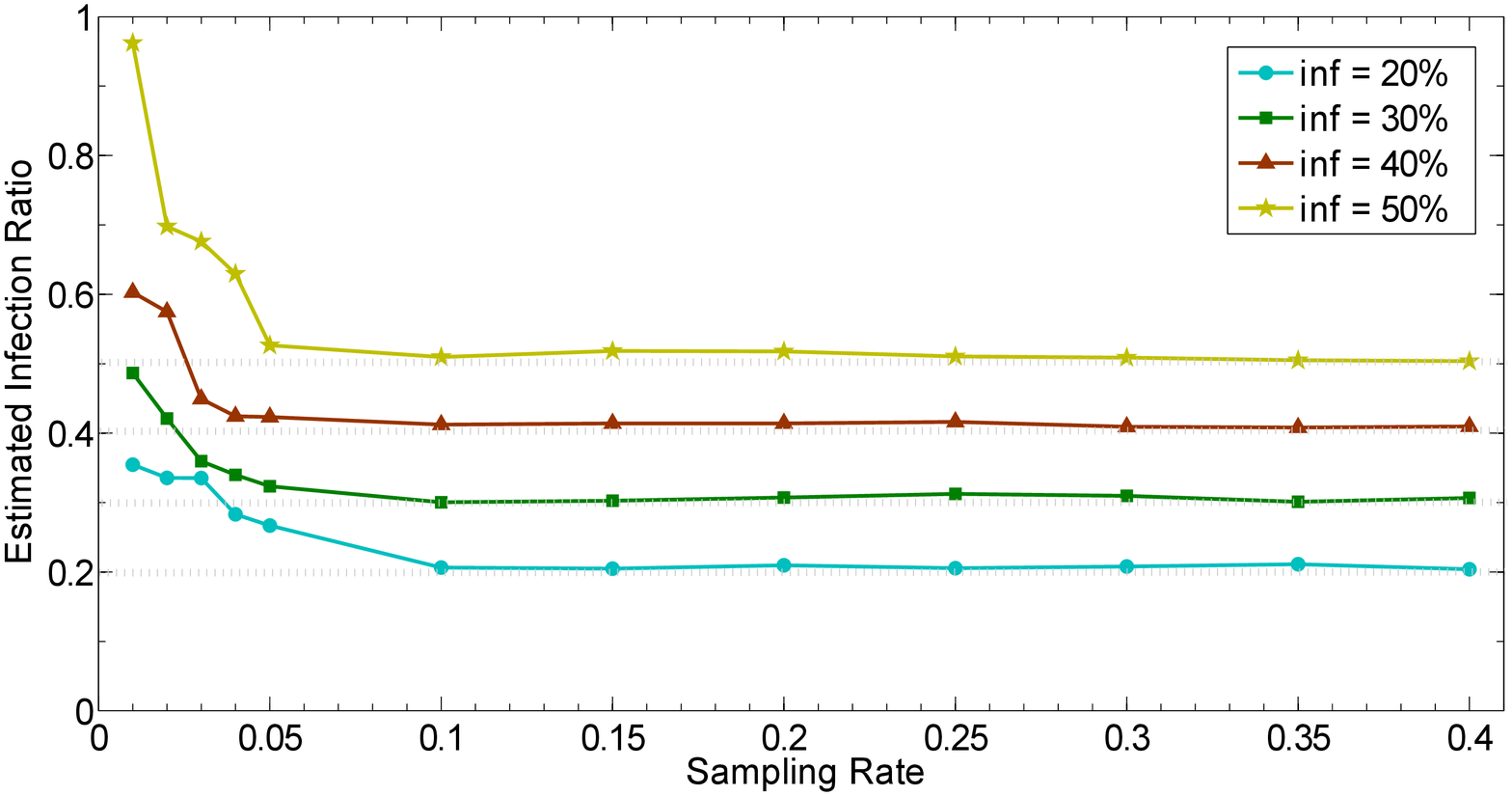}}
    \subfigure[DWS]{\label{DWS-AE}\includegraphics[scale=0.28]{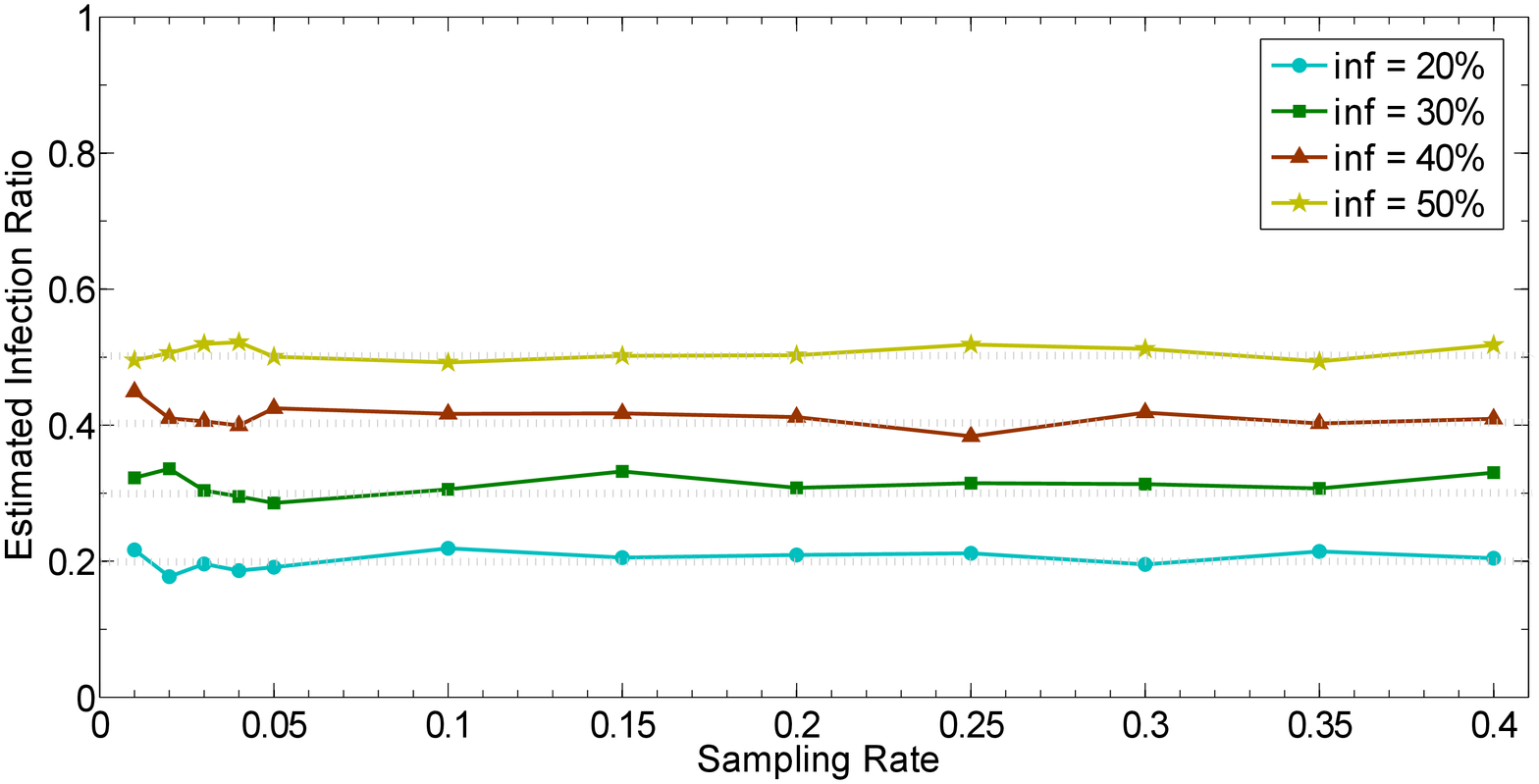}}
    \\
    \subfigure[DSF]{\label{DSF-AE}\includegraphics[scale=0.27]{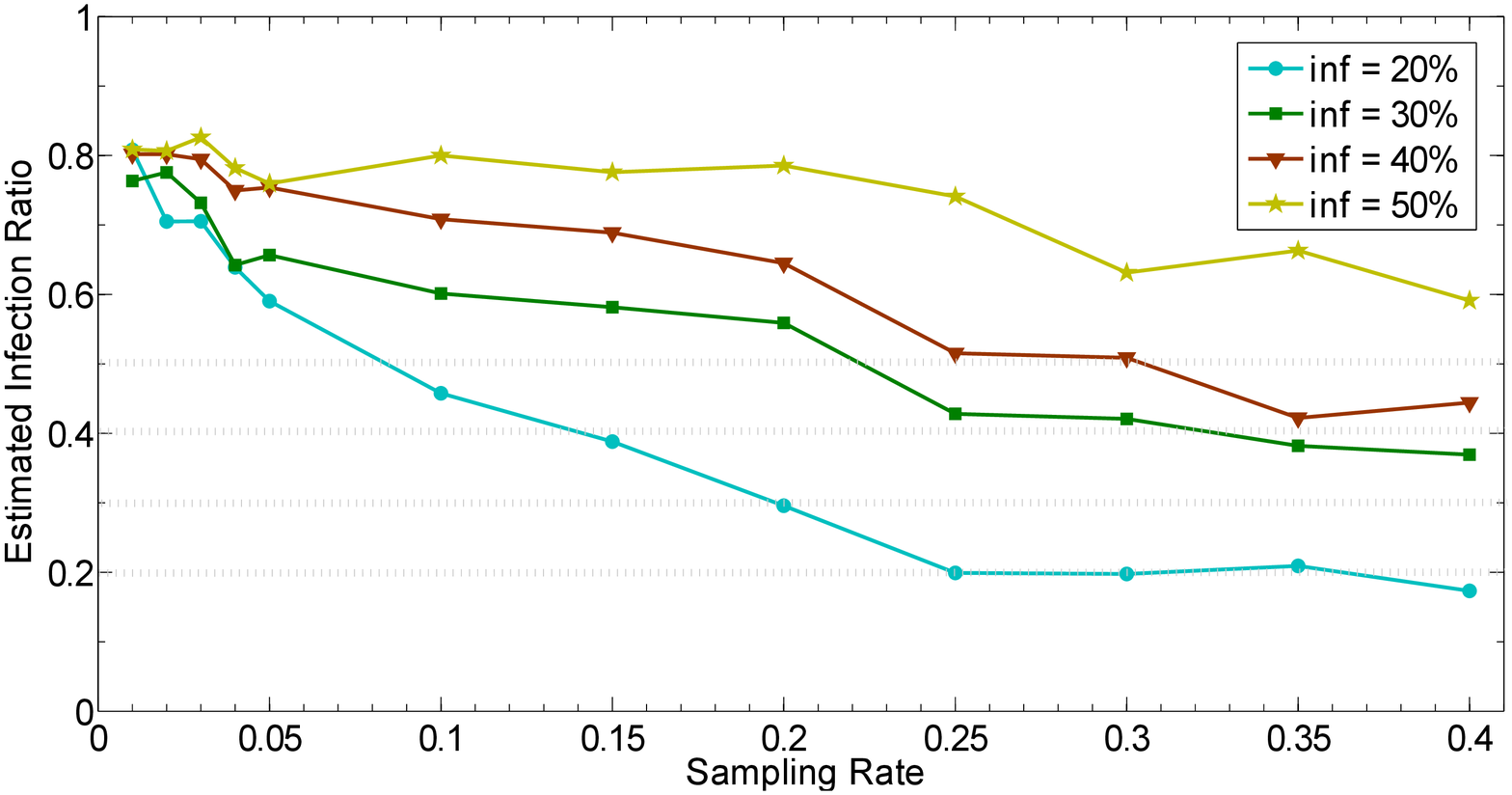}}
	\subfigure[OSN]{\label{OSN-AE}\includegraphics[scale=0.28]{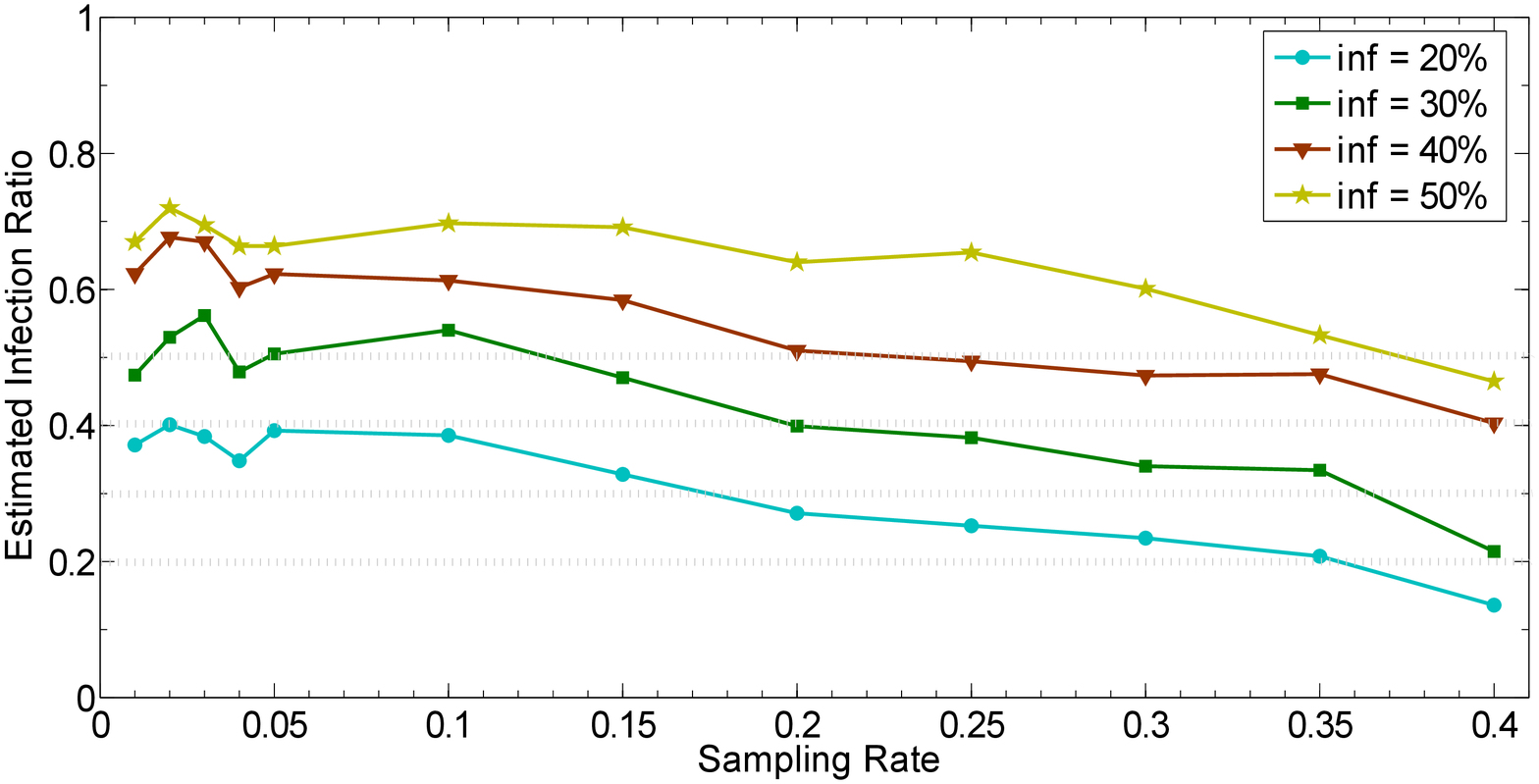}}
 \end{center}
  \caption{The performance of the DNM framework in the networks with different infection ratios; (a) DER, (b) DWS, (c) DSF, and (d) OSN.}
  \label{Inf-whole}
\end{figure*}

Figure \ref{Inf-whole} shows the sensitivity of DNM to infection ratio in DER, DWS, DSF, and OSN datasets (due to page limitation, we did not demonstrate the results for P2P and AS networks). We run the simulation of disease spreading until the true ratio of infected nodes reaches to desired values (specifically, $20\%$, $30\%$, $40\%$, and $50\%$). These ratios are used as the ground truth to performance evaluation. 

According to the results, one can categorize the networks into three groups. The first group includes the networks generated by the DER model. In this group, the estimated infection ratio is more farther from the true value by increasing the true infection ratio. This error converge to zero for higher sampling rate in various infection ratios (In terms of bias, we observe the same pattern; higher infection ratio leads to higher bias in lower sampling rates). In second group, that includes the DWS networks, the infection ratio dose not have major impact on the estimated values and their biases. 

We found an inverse pattern in third group which is composed of the DSF and OSN (we observe the same results in the P2P and AS networks). In particular, lower true infection ratio leads to higher error in estimation of infection ratio (similarly, for obtained bias) for lower sampling rates. However, this error reduces by increasing the number of samples. The deduction rate is higher for the populations with lower infection ratio. As a general conclusion, in a scale free network with higher infection ratio, lower sampling rates are sufficient and result in a reasonable bias. More specifically, increasing the sampling rate does not have significant impact on the bias. 

\subsubsection{Estimating Outdegree Distributions}

In what follows we compare the performance of DNM against the DURW sampling algorithm proposed by Ribeiro et al. \cite{Ribeiro2012} (presented in Section \ref{sec:Related Work}).  
This is the most closely related method to ours that achieves asymptotically unbiased estimates of the outdegree distribution of a directed graph. However, there is a fundamental difference between the proposed framework and DURW; DNM only uses already visited nodes as local information without any prior knowledge about the latent structure of the network. In addition, the DURW algorithm is based on the assumption that nodes can be sampled uniformly at random from the original graph, which is not always feasible. 

\begin{figure}[h]
  \begin{center}
  \includegraphics[scale=0.29]{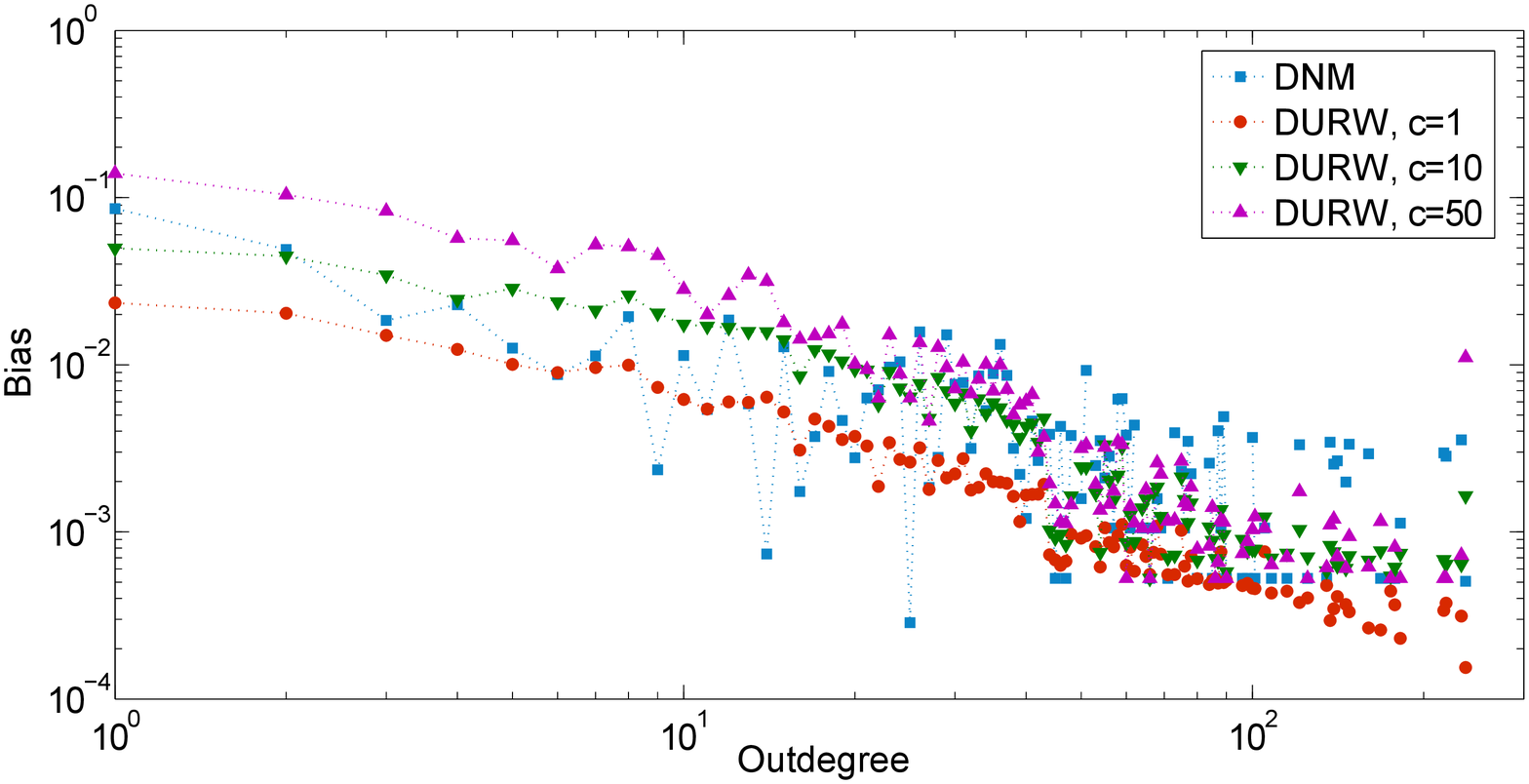}   
\end{center}
  \caption{Bias of DNM compared with DURW ($w =10$ and $c \in \{1, 10, 50\}$) for all outdegrees in OSN dataset  for sampling rate$=10\%$.} 
  \label{fig:outDegree}
\end{figure}

There are two controlling parameters in DURW. The first one is denoted by $c$, which is the cost of a random jump (i.e. the average number of sampling steps required to perform the jump). 
The second one, $w$, is the random jump weight that controls the probability of performing a random jump. 
Figure \ref{fig:outDegree} shows the comparison between estimates of the bias (in log scale) of DNM and DURW for all outdegrees in the OSN dataset (we observed the same results for other test networks). The bias was obtained over 100 runs. In all simulations we sampled $10\%$ of the network. The DURW random jump weight and cost was $w = 10$ and $c \in \{1,10,50\}$, respectively. 

We found that both methods obtain accurate estimates. However, the fraction of nodes with large outdegrees can be estimated more accurately than the ones with small outdegrees. This is because both of them tend to sample nodes with larger outgoing links more frequently; sometimes causing lower estimation errors for the large outdegree nodes. 
Moreover, as demonstrated in Figure \ref{fig:outDegree}, DURW outperforms DNM in terms of bias when the cost of a random jump is not significant. The cost of a jump effectively reduces the number of total nodes that can be sampled, which increases the bias. Therefore, we observe that the bias of the outdegree distribution estimates in DURW increases with $c$, and the DNM method is more accurate than DURW for larger costs.

\section{Conclusion}
\label{sec:Conclusion}
In this paper, we introduced a novel two-step framework to measure nodal characteristics of large scale directed networks which can be defined by average target functions. 
We proposed a novel link-tracing network sampling algorithm by utilizing the idea of personalized PageRank.
In particular, this method samples the underlying network by moving from a node to one of its neighbors through an outgoing link based on the approximated probability of Personalized PageRank. 
Since these probabilities can be considered as an approximation of the exact visiting probability, we proposed a new estimator based on the idea of approximate importance sampling. 
Our estimator is able to overcome the effect of approximate stationary distribution of random walk-based sampling methods on the accuracy of the estimation. We showed both analytically and empirically that the proposed framework is asymptotically unbiased.

\section*{Acknowledgment}
This research has been partially supported by ITRC (Iran Telecommunication Research Center) under grant number 6479/500 (90/4/22).

\end{document}